\def\be{\begin{equation}}
\def\ee{\end{equation}}
\def\ba{\begin{array}}
\def\ea{\end{array}}

\documentclass[11pt]{article}
\usepackage{amsfonts}
\usepackage{amssymb}
\usepackage{dsfont}
\usepackage{amsmath,amssymb,amstext,amsfonts,array,hhline,tabularx,graphics}
\usepackage{amsthm}
\usepackage{multicol}
\usepackage{latexsym}
\usepackage{abstract}
\usepackage{arydshln}
\usepackage{epsfig,graphicx, color}
\usepackage{amsmath}
\usepackage[numbers,sort&compress]{natbib}

\newtheorem{thm}{\bf Theorem}
\newtheorem{defi}{\bf Definition}

\newtheorem{exam}{\bf Example}

\theoremstyle{plain}
\begin{document}
\parskip=3pt
\parindent=18pt
\baselineskip=20pt \setcounter{page}{1}

 \title{\large\bf Separability criteria  based on the correlation tensor moments for arbitrary dimensional states}
\date{}

\author{Xiaofen Huang$^{1, 2}$, Naihuan Jing$^{3}$ \\[10pt]
\footnotesize
\small 1 School of Mathematics and Statistics, Hainan Normal University, Haikou 571158, China\\
\small 2 Key Laboratory of Data Science and Smart Education, \\
\small Ministry of Education, Hainan Normal University,
      Haikou, 571158, China\\
\small 3 Department of Mathematics, North Carolina State University,
Raleigh, NC27695, USA\\       
}
\date{}

\maketitle

\centerline{$^\ast$ Correspondence to  huangxf1206@163.com  }
\bigskip

\begin{onecolabstract}
 \textbf{Abstract:}~~~ As one of the most profound features of quantum mechanics, entanglement is a vital resource for quantum information processing. Inspired by the recent work on PT-moments and separablity [Phys. Rev. Lett. {\bf 127}, 060504 (2021)], we propose two sets of separability criteria using moments of the
 correlation tensor for bipartite and multipartite quantum states, which are shown to be stronger in some aspects in detecting entanglement.

\end{onecolabstract}


\section{Introduction}
Quantum entanglement is a fascinating phenomenon in quantum physics. It is widely recognized as a valuable resource in the rapidly expanding field of quantum information science, with various applications such as quantum algorithms \cite{chuang, algo}, quantum cryptography \cite{cryp1, cryp2}, quanum simulation \cite{simula}, telepotation \cite{telep} and so on.

Detecting wether a quantum state is entangled or not is a fundamental problem of quantum information and quantum computation both in theory and experiments, much efforts have been devoted to its characterization and quantification  \cite{ent1, ent2, ent3, qiao2018, qiao2023}. In \cite{PPT} Peres has given the celebrated PPT criterion which says that for any bipartite separable quantum state the density matrix must be semipositive (PPT) under partial transposition. The PPT is a necessary and sufficient condition of separability for $2\otimes 2$ and $2\otimes 3$ quantum systems \cite{iso}, but not sufficient for high-dimensional states
as the bound entangled states are classes of families of inseparable states with positive partial transposes. Another powerful operational criterion for separability is the realignment criteria \cite{realig}, which can detect entanglement of many bound entangled states.  Recently some further elegant results for the separability problem have been derived \cite{cova, guhne, prl125, prl127, zhang2022}. In \cite{prl125} the authors proposed a separability criterion in terms of moments of the partially transposed density matrix,  which shows that the first three PT-moments can be used to define a simple yet powerful test for bipartite entanglement. Furthermore, some even stronger separability criteria based on the positivity of Hankel matrices involving all PT-moments are presented in \cite{prl127}. The authors in \cite{zhang2022} generalized this method to give separability criteria in terms of  the realignment moments. By the method of matricization of tensors, considerable developments have been made in proposing stronger variants and multipartite generalizations in term of the correlation matrices to detect non-full-separability of multipartite states \cite{cm, vicente2007, vicente2011, sar2020, corr2011, liming, shen2016, wapp1, wapp2, qiao2021}.

The Bloch representation of the density matrix reveals intrinsic property of the quantum state and encodes  its quantum correlation. In \cite{cm, vicente2007} the authors have given a separability criterion based on the correlation matrix of a bipartite state, which seems more efficient than the PPT criterion in many situations.

 This raises a natural question whether more effective tests of entanglement can be found for the quantum state.
It is thus natural to study further properties of the correlation tensor, which
controls the intrinsic properties of the quantum state. In this paper, we propose a refinement method to describe and quantify entanglement. By
 invoking moment vectors in terms of the trace of various powers of the correlation tensor, a stronger separability criterium is derived for  qudits.
The criterion consists of a sequence of moment matrices in terms of correlation tensor moments and their interdependent
relations. If one of the inequalities is violated, then the quantum state is entangled.
We can also use the moment vectors to quantify entanglement.
The new results demonstrate some improvements
in several examples. For instance, our criterion can be used to derive lower bounds for concurrence as an application.

\section{Separability criteria for bipartite states}

Let $\lambda_i^{(d)}, i=1, 2, \cdots, d^2-1$ be the traceless Hermitian generators of $\mathfrak{su}(d)$ satisfying the orthogonality relation ${\rm{Tr}}(\lambda_i^{(d)}\lambda_j^{(d)})=2\delta_{ij}$. Then any state  $\rho$ on $\mathcal{H}_{d_1}\otimes \mathcal{H}_{d_2}$ can be represented as
\begin{equation}\label{decom}
\rho=\frac{1}{d_1d_2}I_{d_1}\otimes I_{d_2}+\sum_{i=1}^{d_1^2-1}r_i\lambda_i^{(d_1)}\otimes I_{d_2}
+\sum_{j=1}^{d_2^2-1}s_j I_{d_1}\otimes \lambda_j^{(d_2)}
+\sum_{i=1}^{d_1^2-1}\sum_{j=1}^{d_2^2-1}T_{ij}\lambda_i^{(d_1)}\otimes \lambda_j^{(d_2)},
\end{equation}
where $I_d$ denotes the $d\times d$ identity matrix, and the coefficients are given by
\begin{equation}
r_i=\frac{1}{2d_2}{\rm{Tr}}\rho\lambda_i^{(d_1)}\otimes I_{d_2},~~~
s_j=\frac{1}{2d_1}{\rm{Tr}}\rho I_{d_1}\otimes \lambda_j^{(d_2)},~~~
T_{ij}=\frac{1}{4}{\rm{Tr}}\rho \lambda_i^{(d_1)}\otimes \lambda_j^{(d_2)}.
\end{equation}
Denote $\textbf{r}=(r_1, r_2, \cdots, r_{d_1^2-1})^t$ and $\textbf{s}=(s_1, s_2, \cdots, s_{d_2^2-1})^t$ as the Bloch vectors in Hilbert spaces $\mathcal{H}_{d_1}$ and $\mathcal{H}_{d_2}$, respectively, where $t$ stands for the transposition. The coefficients $T_{ij}$ form
the {\it correlation matrix} or {\it correlation tensor}
$T=(T_{ij})$ of size $(d_1^2-1)\times (d_2^2-1)$ and the canonical correlation matrix is $\tilde{T}=\begin{pmatrix} \frac{1}{d_1d_2} & \textbf{s}^t \\ \textbf{r} & T \end{pmatrix}$ of size $d_1^2\times d_2^2$.

In the following we present separability criteria based on the moments of the (canonical) correlation tensors using the H\"older inequality and Schatten-$p$ norm of matrices. First of all, we define the moments of the correlation and canonical correlation tensors of the given quantum state $\rho$.
\begin{defi}
The moments of correlation tensor and the canonical correlation tensor are defined by
\[
\begin{split}
&a_k: ={\rm{Tr}}(TT^{\dag})^{\frac{k}{2}}, ~~k=0, 1, \cdots, (d_1^2-1) (d_2^2-1),\\
&b_l: ={\rm{Tr}}(\tilde{T}\tilde{T}^{\dag})^{\frac{l}{2}}, ~~l=0, 1, \cdots, d_1^2 d_2^2.
\end{split}
\]
\end{defi}

For convenience we set $a_0=(d_1^2-1)(d_2^2-1)$, $b_0=d_1^2d_2^2$. Denote the correlation tensor moment vector and the canonical correlation tensor moment vector as $\textbf{a}=(a_0, a_1, \cdots, a_{(d_1^2-1)(d_2^2-1)})$ and
$\textbf{b}=(b_0, b_1, \cdots, b_{ d_1^2 d_2^2})$,  respectively.
Then we have the following result on quantum separability in terms of the (canonical) correlation tensor moments.

\begin{thm}\label{thm1}
If a quantum state $\rho$ in bipartite system $\mathcal{H}_{d_1}\otimes \mathcal{H}_{d_2}$ is separable, then the following inequalities hold
\begin{equation}\label{bi1}
a_2^2\leq \frac{\sqrt{(d_1^2-d_1)(d_2^2-d_2)}}{2d_1d_2}a_3,
\end{equation}
\begin{equation}\label{bi2}
b_2^2\leq\frac{\sqrt{(2+d_1^2-d_1)(2+d_2^2-d_2)}}{2d_1d_2}b_3.
\end{equation}
\end{thm}

\textbf{Proof}
The H$\rm{\ddot{o}}$lder inequality for $n$-dimensional vectors $v=(v_i)$ and $w=(w_i)$ says that
\begin{equation}
|\langle v, w\rangle|=\sum_{i=1}^n v_iw_i\leq \| v\|_{l_p}\|w\|_{l_q},
\end{equation}
where $p, q\geq 1$,  $\frac{1}{p}+\frac{1}{q}=1$,  and $\|v\|_{l_p}:=(\sum_i|v_i|^p)^{\frac{1}{p}}$ is the $l_p$ norm of $v$.
The Schatten-$p$ norm of Hermitian matrix $M$ is given by
\begin{equation}
\|M\|_p=\Big({\rm{Tr}}(MM^{\dag})^{\frac{p}{2}}\Big)^{\frac{1}{p}}=(\sum_i\sigma_i^p)^{\frac{1}{p}},
\end{equation}
where $\sigma_i$'s are the singular values of matrix $M$. Note that $\|M\|_1$ reduces to the trace norm $\|M\|_{tr}$ of $M$. Also
$\|M\|_2^4=(\sum_{i}\sigma_i^2)^2$, $\|M\|_1=\sum_i\sigma_i$, and $\|M\|_3^3=\sum_i\sigma_i^3$. Let $v_i=\sigma_i^{\frac{1}{2}}$ and $w_i=\sigma_i^{\frac{3}{2}}$ in the H\"older inequality, then
\begin{equation}
\sum_{i}\sigma_i^2=\sum_i \sigma_i^{\frac{1}{2}}\sigma_i^{\frac{3}{2}}\leq (\sum_i\sigma_i)^{\frac{1}{2}}(\sum_i\sigma_i^3)^{\frac{1}{2}},
\end{equation}
i. e.,
\begin{equation}
(\sum_{i}\sigma_i^2)^2\leq (\sum_i\sigma_i)(\sum_i\sigma_i^3).
\end{equation}
Therefore, for any Hermitian matrix $M$ one obtains
\begin{equation}\label{bi3}
\|M\|_2^4\leq \|M\|_1 \|M\|_3^3.
\end{equation}

Now take $M=(TT^{\dag})^{\frac{1}{2}}$, where $T$ is the correlation matrix of $\rho$ in the Bloch decomposition (\ref{decom}). Hence
$\|M\|_2^4=a_2^2$, $\|M\|_1=a_1$ and $\|M\|_3^3=a_3$. By inequality (\ref{bi3}), for any bipartite state $\rho$ we have
\begin{equation}\label{bi33}
a_2^2\leq a_1a_3.
\end{equation}

 If $\rho$ is separable, the correlation matrix criterion in \cite{vicente2007} implies that the trace norm satisfies $\|T\|_{tr}={\rm {Tr}} (TT^{\dag})^\frac{1}{2}=a_1\leq \frac{\sqrt{(d_1^2-d_1)(d_2^2-d_2)}}{2d_1d_2}$. Substituting into the inequality (\ref{bi33}), one obtains the inequality (\ref{bi1}).

 Furthermore, for $M=(\tilde{T}\tilde{T}^{\dag})^{\frac{1}{2}}$, it has been shown that the canonical correlation matrix $\tilde{T}$ satisfies relation
$\|\tilde{T}\|_{tr}={\rm {Tr}} (\tilde{T}\tilde{T}^{\dag})^\frac{1}{2}=b_1\leq \frac{\sqrt{(2+d_1^2-d_1)(2+d_2^2-d_2)}}{2d_1d_2}$ for any bipartite separable state \cite{liming}. Thus the inequality (\ref{bi2}) can be proved similarly. $\Box$

The separability criteria in Theorem \ref{thm1} use the first three (canonical) correlation tensor moments.
In the following improved separability criteria involving higher order (canonical) correlation tensor moments are derived.

First of all,  we need to construct two families of semipositive Hankel matrices (cf. \cite{prl127}), 
i. e.,  matrices $H_k(\textbf{a})$ ($H_k(\textbf{b})$) with entries $[H_k(\textbf{a})]_{ij}=a_{i+j}$ ($[H_k(\textbf{b})]_{ij}=b_{i+j}$) for $k=1, 2, \cdots, \lfloor \frac{d_1d_2}{2}\rfloor $ and $i, j=0, 1, 2, \cdots, k$,   and $B_l(\textbf{a})$ ($B_l(\textbf{b})$) with entries  $[B_l(\textbf{a})]_{mn}=a_{m+n+1}$ ($[B_l(\textbf{b})]_{mn}=b_{m+n+1}$) for $l=1, 2, \cdots, \lfloor \frac{d_1d_2-1}{2}\rfloor$ and $m, n=0, 1, 2, \cdots, l$, where $\lfloor \cdot \rfloor$ stands for the integer function. Then we can propose separability criteria based on the Hankel matrices.

\begin{thm}\label{thm2}
Given a separable bipartite state $\rho$  in Hilbert space $\mathcal{H}_{d_1}\otimes \mathcal{H}_{d_2}$, then the Hankel matrices satisfy conditions below
\begin{equation}\label{bi4}
\widehat{H_k}(\textbf{a})\geq 0,~~ \widehat{B_l}(\textbf{a})\geq 0,
\end{equation}
and
\begin{equation}\label{bi5}
\widehat{H_k}(\textbf{b})\geq 0, ~~\widehat{B_l}(\textbf{b})\geq 0.
\end{equation}
where $\widehat{H_k}(\textbf{a})$ $(resp. \widehat{H_k}(\textbf{b}))$ and $\widehat{B_l}(\textbf{a})$ $(resp. \widehat{B_l}(\textbf{b}))$ are the matrices obtained by replacing $a_1= \frac{\sqrt{(d_1^2-d_1)(d_2^2-d_2)}}{2d_1d_2}$ $(resp. b_1=\frac{\sqrt{(2+d_1^2-d_1)(2+d_2^2-d_2)}}{2d_1d_2})$ in the Hankel matrices  $H_k(\textbf{a})$ $(resp. H_k(\textbf{b}))$ and
$B_l(\textbf{a})$ $(resp. B_l(\textbf{b}))$.
\end{thm}

\textbf{Proof} For any bipartite state $\rho$ with the Bloch decomposition as Eq. (\ref{decom}), set $H=(TT^{\dag})^{\frac{1}{2}}$. We introduce two matrix vectors
\[
 \textbf{x}:=(x_0, x_1, \cdots, x_{\lfloor \frac{d_1d_2}{2}\rfloor})
 = (I=H^0, H, H^2,  \cdots, H^{\lfloor \frac{d_1d_2}{2}\rfloor}),
\]
and
\[
\textbf{y}:=(y_0, y_1, \cdots, y_{\lfloor \frac{d_1d_2-1}{2}\rfloor})
= (H^{\frac{1}{2}}, H^{\frac{3}{2}}, \cdots, H^{\lfloor \frac{d_1d_2-1}{2}\rfloor}).
\]

Under the Hilbert-Schmidt inner product of matrices, we can construct two Gram matrices in terms of $\textbf{x}$ and $\textbf{y}$ 
by
$\langle x_i, x_j\rangle={\rm{Tr}} H^{i+j}=a_{i+j}$ and $ \langle y_i, y_j\rangle={\rm{Tr}} H^{i+\frac{1}{2}}H^{j+\frac{1}{2}}={\rm{Tr}} H^{i+j+1}=a_{i+j+1}$.
So the Gram matrices happen to be the Hankel matrices $H_k(\textbf{a})$ and $B_l(\textbf{a})$, respectively. It is known that the Gram matrices are always positive semidefinite, thus $H_k(\textbf{a})\geq 0$ and $B_l(\textbf{a})\geq 0$ for any state.

Since the first correlation tensor moment $a_1=\|T\|_{tr} \leq \frac{\sqrt{(d_1^2-d_1)(d_2^2-d_2)}}{2d_1d_2}$ for any separable bipartite $\rho$ \cite{vicente2007}.  Replacing all $a_1$'s in $H_k(\textbf{a})$ and  $B_l(\textbf{a})$ by $\frac{\sqrt{(d_1^2-d_1)(d_2^2-d_2)}}{2d_1d_2}$, we obtain the inequalities (\ref{bi4}).

Similarly, if we let $H=(\tilde{T}\tilde{T}^{\dag})^{\frac{1}{2}}$,  then $b_1=\|\tilde{T}\|_{tr}\leq \frac{\sqrt{(2+d_1^2-d_1)(2+d_2^2-d_2)}}{2d_1d_2}$ for a separable bipartite $\rho$ \cite{liming}.  Replace  all $b_1$'s in $B_l(\textbf{a})$ by $\frac{\sqrt{(2+d_1^2-d_1)(2+d_2^2-d_2)}}{2d_1d_2}$, then we get the inequalities (\ref{bi5}) in the Theorem. This completes the proof.  $\Box$

Note that Theorem \ref{thm1} is a special case of Theorem \ref{thm2}. If we choose $a_1=\frac{\sqrt{(d_1^2-d_1)(d_2^2-d_2)}}{2d_1d_2}$ and $b_1=\frac{\sqrt{(2+d_1^2-d_1)(2+d_2^2-d_2)}}{2d_1d_2}$, by Theorem \ref{thm2}, the Hankel matrices $\widehat{B_1}(\textbf{a})=\begin{pmatrix} a_1 & a_2\\ a_2 & a_3 \end{pmatrix}\geq 0$ and $\widehat{B_1}(\textbf{b})=\begin{pmatrix} b_1 & b_2\\ b_2 & b_3 \end{pmatrix}\geq 0$, which reduce to the inequalities in Theorem \ref{thm1}.

Obviously, the equalities in Theorem 1 and Theorem 2 hold for the separable state $\frac{1}{d_1d_2}I$, as in this case all the moments are zero.
We remark that the correlation tensor moments $a_i$'s and $b_i$'s  are closely related with the $i$th powers of singular values of the correlation matrix.
In fact, a fundamental theorem of algebra says that all roots of polynomial are determined by their coefficients, which implies that
 the moments actually completely determine the singular values of the correlation tensor.
 Therefore the criteria in Theorems \ref{thm1} and \ref{thm2} are intrinsically more refined to measure the entanglement of quantum states. Furthermore, correlation tensor moments can be estimated by experiments thus might be used in detecting
 entanglement in experiments.

\section{Separability criteria for multipartite states}

How to quantify entanglement of multipartite quantum states is a fundamental problem in quantum information. In this section, we derive a criterion of fully separability based on the (canonical) correlation tensor moments for multipartite quantum states.

Let $\lambda_{\alpha_k}^{\{\mu_k\}}=I_{d_1}\otimes \cdots \otimes I_{d_{\mu_{k-1}}}\otimes\lambda_{\alpha_k} \otimes I_{d_{\mu_{k+1}}}\otimes \cdots \otimes I_{d_n}$ with $\lambda_{\alpha_k}$, the generators of $\mathfrak{su}(d_i)$, appearing at the $\mu_k$th position. For a multipartite state $\rho$, let
\begin{equation}\label{tensor}
\mathcal{T}_{\alpha_1\alpha_2\cdots \alpha_m}^{\mu_1\mu_2\cdots \mu_m}=\frac{\prod_{i=1}^m d_{\mu_i}}{2^m \prod_{i=1}^n d_i}{\rm {Tr}}\rho \lambda_{\alpha_1}^{\{\mu_1\}}\lambda_{\alpha_2}^{\{\mu_2\}}\cdots \lambda_{\alpha_m}^{\{\mu_m\}},
\end{equation}
be the entries of the tensor $\mathcal{T}^{\{\mu_1\mu_2\cdots \mu_m\}}$.

For $\alpha_m=\cdots= \alpha_n=0$ with $1\leq m\leq n$, we define that $\mathcal{\tilde{T}}_{\alpha_1\alpha_2\cdots\alpha_n}=\mathcal{T}_{\alpha_1\cdots \alpha_m}^{\mu_1\cdots \mu_m}$, and for $\alpha_1=\cdots =\alpha_n=0$, define that $\mathcal{\tilde{T}}_{\alpha_1\alpha_2\cdots\alpha_n}=\frac{1}{\prod_{k=1}^nd_k}$. For generally, let $\lambda_{\alpha_0}^{\mu_k}=I_{d_i}$ in
\eqref{tensor}, we then define the extended tensor
$\mathcal{\tilde{T}}$ with entries $\mathcal{\tilde{T}}_{\alpha_1\alpha_2\cdots\alpha_n}, \alpha_k=0, 1, \cdots, d_k^2-1$.

If we set $\lambda_0^{\{k\}}=I_{d_k}$ for any $1\leq k\leq n$, then any multipartite state $\rho\in\mathcal{H}_{d_1}\otimes \mathcal{H}_{d_2}\otimes \cdots \otimes \mathcal{H}_{d_n}$ can be generally expressed by the tensor $\mathcal{\tilde{T}}$ as
\begin{equation}\label{multi}
\rho=\sum_{\alpha_1\alpha_2\cdots \alpha_n}\mathcal{\tilde{T}}_{\alpha_1\alpha_2\cdots \alpha_n}\lambda_{\alpha_1}^{\{1\}}\lambda_{\alpha_2}^{\{2\}}\cdots \lambda_{\alpha_n}^{\{n\}},
\end{equation}
where the summation is taken for all $a_k=0, 1, \cdots, d_k^2-1$.
Next we adopt the definition of the $k$th matrix unfolding $\mathcal{T}_k$ of a tensor $\mathcal{T}$, which is a matrix with $i_k$ to be the row index and the rest subscripts of $\mathcal{T}$ to be column indices (detailed description can be found in Ref. \cite{hassan}).
The Schatten-$p$ norm of the tensor $\mathcal{T}$ over $n$ matrix unfoldings is defined as
\begin{equation}
\|\mathcal{T}\|_p:=max\{\|\mathcal{T}_k \|_p\}, k=1, 2, \cdots, n.
\end{equation}

To obtain the separability criterion for $n$-partite quantum systems, we introduce the (canonical) correlation tensor moments for multipartite quantum states.
\begin{defi} Let $\rho$ be a multipartite state in Hilbert space $\mathcal{H}_{d_1}\otimes \mathcal{H}_{d_2}\otimes \cdots \otimes \mathcal{H}_{d_n}$ with decomposition as Eq. (\ref{multi}),  the (canonical) correlation tensor moments are given by
\[
\begin{split}
&\bar{a}_i:={\rm{Tr}}(\mathcal{T}\mathcal{T}^{\dag})^{\frac{i}{2}},  i=1, 2, \cdots, \prod_{k=1}^n(d_k^2-1), \\
&\bar{b}_j:={\rm{Tr}}(\mathcal{\tilde{T}}\mathcal{\tilde{T}}^{\dag})^{\frac{j}{2}},  j=1, 2, \cdots, \prod_{k=1}^n d_k^2.
\end{split}
\]
\end{defi}

Recall that a multipartite quantum state in $\mathcal{H}_{d_1}\otimes \mathcal{H}_{d_2}\otimes \cdots \otimes \mathcal{H}_{d_n}$ is fully separable \cite{hassan} if and only if it can be written as a convex sum of tensor products of subsystem states
\begin{equation}
\rho=\sum_ip_i\rho_i^{(d_1)}\otimes \cdots \otimes \rho_i^{(d_n)},
\end{equation}
where the probabilities $p_i\geq 0$, $\sum_ip_i=1$, and $\rho_i^{(d_1)}, \cdots, \rho_i^{(d_n)}$ are pure states of the subsystems. A state is called $k-$partite separable if we can write
\begin{equation}
\rho=\sum_{i}p_i\rho_{i}^{(a_1)}\otimes \rho_{i}^{(a_2)}\cdots \otimes \rho_{i}^{(a_k)},
\end{equation}
where $a_i$, $i=1, 2, \cdots, k$ are the disjoint subsets of $\{1, 2, \cdots, n\}$ and $\rho^{(a_i)}$ acts on the tensor product space made up by the factors of $\mathcal{H}$ labeled by the members of $a_i$.

If a state is separable for a bipartite partition, it is called biseparable. And a state is genuine entanglement if it is not separable for any partition.

In \cite{hassan, liming}, the authors have given the generalized forms of separability criteria based on the
correlation tensor and canonical correlation tensor which says that if a quantum state $\rho \in \mathcal{H}_{d_1}\otimes \mathcal{H}_{d_2}\otimes \cdots \otimes \mathcal{H}_{d_n}$
is fully separable, then
\begin{equation}\label{gene}
\|\mathcal{T}\|_{tr}\leq \prod _{k=1}^n \sqrt{\frac{d_k-1}{2d_k}},~~\|\tilde{\mathcal{T}}\|_{tr}\leq\prod _{k=1}^n\sqrt{\frac{d_k^2-d_k+2}{2d_k^2}}.
\end{equation}
In the following we show that one can obtain the generalized correlation tensor moment criterion from Theorem \ref{thm1}.
\begin{thm}\label{thm3}
If a quantum state $\rho \in \mathcal{H}_{d_1}\otimes \mathcal{H}_{d_2}\otimes \cdots \otimes \mathcal{H}_{d_n}$ is fully separable, then the following inequalities must hold
\begin{equation}
\bar{a}_2^2\leq \bar{a}_3\prod _{k=1}^n \sqrt{\frac{d_k-1}{2d_k}},
\end{equation}
\begin{equation}
\bar{b}_2^2\leq \bar{b}_3 \prod _{k=1}^n\sqrt{\frac{d_k^2-d_k+2}{2d_k^2}}.
\end{equation}
In other words, if one of the inequalities is violated, then the state $\rho$ is entangled.
\end{thm}

\textbf{Proof} Note that $\bar{a}_1=\|\mathcal{T}\|_{tr}$ and $\bar{b}_1=\|\tilde{\mathcal{T}}\|_{tr}$, then the inequalities are similarly shown as
that of Theorem \ref{thm1}.

As we have remarked that the singular values of correlation matrix reveal intrinsic information about entanglement of the quantum state,
the moment vectors provide overall information of the singular values and easier to compute.
Thus Theorem 3 gives a refined and balanced method to detect entanglement, while Ref.\cite{liming} involves only the first power of singular values of the correlation matrix.

{\section{Efficiency of the Criterion}}

{The following examples illustrate feasibility and effectiveness of our separability
criteria in detecting entanglement for the Werner states and the PPT entangled states. They show that the criteria provide some general operational and
 sufficient conditions to detect entanglement in quantum states of arbitrary dimensions.}

 {
\begin{exam}
The Werner states in bipartite system $\mathcal{H}_d\otimes \mathcal{H}_d$  can be represented as
\begin{equation}
\rho_W=\frac{1}{d^3-d}[(d-x)I\otimes I+(dx-1)F],
\end{equation}
where $-1\leq x\leq 1$, $F$ is the flip operator defined by $F(\phi\otimes \varphi)=\varphi\otimes \phi$. These states are separable iff $x\geq 0$ \cite{werner} and the Bloch representation is given by
\begin{equation}
\rho_W=\frac{1}{d^2}(I\otimes I+\frac{d(dx-1)}{2(d^2-1)}\lambda_i\otimes \lambda_i).
\end{equation}
Theorem \ref{thm1} recognizes separability when $\frac{2-d}{d}\leq x\leq 1$, while the separability criterion in \cite[Prop. 3]{vicente2007} detects separability for
$ \frac{d-2}{d(d-1)}\leq x\leq \frac{1}{d-1}$. Clearly our criterion is stronger than that of \cite{vicente2007}.
\end{exam}}


\begin{exam}
Consider the $3\times 3$ PPT entangled state \cite{cryp2}:
\begin{equation}\label{e3}
\rho=\frac{1}{4}(I_9-\sum_{i=0}^4|\chi_i\rangle\langle\chi_i|),
\end{equation}
where $|\chi_0\rangle=|0\rangle(|0\rangle-|1\rangle)/\sqrt{2}$, $|\chi_1\rangle=(|0\rangle-|1\rangle)|2\rangle/\sqrt{2}$, $|\chi_2\rangle=|2\rangle(|1\rangle-|2\rangle)/\sqrt{2}$, $|\chi_3\rangle=(|1\rangle-|2\rangle)|0\rangle/\sqrt{2}$, $|\chi_4\rangle=(|0\rangle+|1\rangle+|2\rangle)(|0\rangle+|1\rangle+|2\rangle)/3$.
Let $\rho$ be the mixed state with white noise:
\[
{\rho_x}=x\rho+\frac{1-x}{9}I_9, ~~~0 \leq x\leq 1.
\]
The correlation matrix criterion \cite{vicente2007} shows that {$\rho_x$} is entangled for $0.9493<x\leq 1$, and the separability criterion \cite{liming}
says that it is entangled when $0.89254< x\leq 1$. Using the separability criterion (\ref{bi1}) in Theorem \ref{thm1}, one can show that {$\rho_x$} is entangled for $0.84327\leq x\leq 1$,
which is stronger than both tests.
\end{exam}

It is  noted that that the examples above show that our criteria provide an operational method on the sufficient conditions for entanglement, since
the inequalities in Theorem \ref{thm3} provide the least set of constraints for the separability and
confinement of the singular values. As the moments represent global information on entanglement, while singular values reveal the local information,
it is easier to verify. In other words, the inequalities Theorem $\ref{thm1}$, Theorem $\ref{thm2}$ and Theorem $\ref{thm3}$ of the moments $a_i$'s give information about the lower bound for entanglement.
By comparing lower bounds for entanglement of various criteria, it shows that our criteria is stronger than that of criterion in \cite{vicente2007} and \cite{liming}.


\section{Conclusions}

We have investigated the entanglement detection for both bipartite and multipartite quantum systems.
Using the correlation tensors of the Bloch representation, we have provided a necessary condition of separability based on the two types of
tensor moments. Detailed examples show that our separability criteria are more effective than  the criteria established before based on the correlation matrices such as Li'criterion \cite{liming} and dV'criterion \cite{vicente2011}. Our method has used the Bloch representation
in term of the usual Gell-Mann type generators of $\mathfrak{su}(d)$, it would be interesting to explore the Bloch representation using other types of basis
such the Weyl operators \cite{huang2022}, which may lead to other effective separability criterion.
Since it is widely believed that a stronger violation of a separability condition indicates a bigger amount of entanglement, it has been shown that they can be used to place lower bounds on different entanglement measures \cite{julio}. As an application, our criterion can be used to derive lower bounds for concurrence.

\bigskip
\noindent{\bf Acknowledgments}

The research is partially supported by 
the Simons Foundation
(grant 563828) and the specific research fund of the Innovation Platform for Academicians of Hainan Province under Grant No. YSPTZX202215.

\textbf{Date Availability Statement}
All data generated or analysed during this study are available from the corresponding author on reasonable request.

\textbf{Conflict of Interest Statement}
We declare that we have no conflict of interest.


\begin{thebibliography}{00}
\bibitem{chuang}M. A. Nielsen and I. L. Chuang, Quantum computation and quantum information (Cambridge University Press, Cambridge, England, 2000).
\bibitem{algo}D. Bruss and C. Macchiavello, Multipartite entanglement in quantum algorithms, Phys. Rev. A {\bf 83}, 052313 (2011).
\bibitem{cryp1}A. K. Ekert, Quantum cryptography based on Bell's theorem, Phys. Rev. Lett. {\bf 67}, 661 (1991).
\bibitem{cryp2}S. Schauer, M. Huber and B. C. Hiesmayr, Experimentally feasible security check for n-qubit quantum secret sharing, Phys. Rev. A {\bf 82}, 062311 (2010).
\bibitem{simula}S. Lioyd, Universal quantum simulators, Science {\bf 273}, 1073 (1996).
\bibitem{telep}T. Gao, F. L. Yan and Y. C. Li, Optimal controlled teleportation, Europhys. Lett. {\bf 84}, 50001 (2008).
\bibitem{ent1}L. Zhang, Z. Chen, S.-M. Fei, Entanglement verification with deep semi-supervised machine learning, Phys. Rev. A {\bf 108}, 022427(2023).
\bibitem{ent2}X. Zhu, J. Wang, G. Bao, M. Li, S.-Q. Shen, S.-M. Fei, A family of bipartite separability criteria based on Bloch representation of density matrices, Quant. Inf. Process. {\bf 22}, 185(2023).
\bibitem{ent3}Y. Li,J. Shang, Geometric mean of bipartite concurrences as a genuine multipartite entanglement measure, Phys. Rev. Research {\bf 4}, 023059(2022).
\bibitem{qiao2018}S. M. Zangi, J.-L. Li, and C.-F. Qiao, Quantum state concentration and classification of multipartite entanglement, Phys. Rev. A {\bf97}, 012301 (2018).
\bibitem{qiao2023}A. Rahman, H. Ali, S. Haddadi , S.M. Zangi, Generating non-classical correlations in two-level atoms, Alexandria Engineering Journal {\bf67}, 425(2023).
\bibitem{PPT}A. Peres, Separability criterion for density matrices, Phys. Rev. Lett. {\bf 77}, 1413 (1996).
\bibitem{iso}M. Horodecki, P. Horodecki and R. Horodecki, Separability of mixed states: necessary and sufficient conditions, Phys. Lett. A {\bf 223}, 1 (1996).
\bibitem{realig}K. Chen, L. A. Wu, A matrix realignment method for recognizing entanglement, Quantum Inf. Comput. {\bf 3}, 193 (2002).
\bibitem{cova}O. G\"uhne, P. Hyllus, O. Gittsovich and J. Eisert, Covariance matrices and the separability problem, Phys. Rev. Lett. {\bf 99}, 130504 (2007).
\bibitem{guhne}O. G\"uhne and G. T\'oth, Entanglement detection, Phys. Rep. {\bf 474}, 1(2009).
\bibitem{prl125}A. Elben, R. Kueng, H. Huang, R. Bijnen, C. Kokail, M. Dalmonte, P. Calabrese, B. Kraus, J. Preskill, P. Zoller and B. Vermersch, Mixed-state entanglement from local randomized measurements, Phys. Rev. Lett. {\bf 125}, 200501 (2020).

\bibitem{prl127}X. Yu, S. Imai and O. G\"uhne, Optimal entanglement certification from moments of the partial transpose, Phys. Rev. Lett. {\bf 127}, 060504 (2021).
\bibitem{zhang2022}T. Zhang, N. Jing and S.-M. Fei, Quantum separability criteria based on realignment moments, Quant. Inf. Process. {\bf 21}, 276 (2022).

\bibitem{cm}J. D. Vicente, Further results on entanglement detection and quantification from the correlation matrix criterion, J. Phys. A: Math. Theor. {\bf 41}, 065309 (2008).
\bibitem{vicente2007}J. I. de Vicente, Separability criteria based on the Bloch representation of density matrices, Quantum Inf. Comput. {\bf 7}, 624 (2007).
\bibitem{vicente2011}J. I. de Vicente and M. Huber, Multipartite entanglement detection from correlation tensors, Phys. Rev. A {\bf 84}, 062306 (2011).
\bibitem{sar2020}G. Sarbicki, G. Scala and D. Chru\'sci\'nski, Family of multipartite separability criteria based on a correlation tensor,
 Phys. Rev. A {\bf 101}, 012341 (2020).
 \bibitem{corr2011}W. Laskowski, M. Markiewicz, T. Paterek and M. \.Zukowski, Correlation-tensor criteria for genuine multiqubit entanglement, Phys. Rev. A {\bf 84}, 062305 (2014).
\bibitem{liming}M. Li, J. Wang, S.-M. Fei and X. Li-Jost, Quantum separability criteria for arbitrary-dimensional multipartite states. Phys. Rev. A {\bf 89}, 022325 (2014).
\bibitem{shen2016}S. Shen, J. Yu, M. Li and S.-M. Fei, Improved separability criteria based on Bloch representation of density matrices, Sci. Rep.  {\bf 6}, 28850 (2016).
\bibitem{wapp1}J. Chang, M. Cui, T. Zhang and S.-M. Fei, Separability criteria based on Heisenberg-Weyl representation of
density matrices, Chin. Phys. B {\bf 27}, 030302 (2018).
\bibitem{wapp2}H. Zhao, Y. Yang, N. Jing, Z. X. Wang and S.-M. Fei, Detection of multipartite entanglement based on Heisenberg-Weyl representation of
density matrices, Quantum Inf. Process. {\bf 19}, 1(2020).
\bibitem{qiao2021}S. M. Zangi, C.-F.Qiao, Robustness of entangled states against qubit loss, Phys. Lett. A {\bf400}, 127322 (2021).
\bibitem{hassan}A. S. M. Hassan and P. S. Joag, Separability criterion for multipartite quantum states based on the Bloch representation of density matrices, Quantum. Inf. Comput. {\bf 8}, 773 (2008).

\bibitem{werner}R. F. Werner, Quantum states with Einstein-Podolsky-Rosen correlations admitting a hidden-variable model, Phys. Rev. A {\bf 40}, 4277 (1989).
\bibitem{huang2022} X. Huang, T. Zhang, M. Zhao and N. Jing, Separability criteria based on the Weyl operators, Entropy {\bf 24}, 1064 (2022).


\bibitem{julio}J. I. Vicente, Further results on entanglement detection and quantification from the correlation matrix criterion, J. Phys. A: Math. Theor.{\bf 41}, 065309(2008).

\end{thebibliography}
\end{document}